# Investigation of the $^{232}$Th Nucleus Excitations at the FEL γ - Nucleus Colliders

H. Koru [a], A. Özcan [a], S. Sultansoy [a,b] and B. Şarer [a]

[a] Physics Dept., Faculty af Science and Arts, Gazi University, Ankara, TURKEY
[b] Institute of Physics, Academy of Sciences, Baku, AZERBAIJAN

## Abstract

The physics search potential of the FEL γ - Nucleus colliders is analyzed using excitations of the $^{232}$Th nucleus. It is shown that, due to the monochromacity of FEL γ beam and high statistics, proposed colliders will play an important role in the field of "traditional" nuclear physics.

## 1. Introduction

An investigation of nucleus excitations still stay as powerful way to clarify different aspects of nuclear physics. There are a lot of different methods to produce excited nucleus: (p, t), (p, p'), (d, d'), (n, γ), (γ, γ'), (e, e'), (n, α) etc [1].

Recently, the new method, namely FEL γ - Nucleus colliders, have been proposed in [2, 3]. In this study we demonstrate the capacity of these machines using Th nucleus as an example.

## 2. FEL γ - Nucleus colliders

The main idea is very simple [2]: relativistic ions will "see" FEL beam as a "laser" photons with energy $\omega = 2\gamma_A \omega_0$, where $\gamma_A$ is the Lorentz factor of ions and $\omega_0$ is the energy of the FEL photons. Therefore, keV energy FEL will give opportunity to investigate MeV energy excitations of relativistic nuclei. Moreover, since the accelerated nuclei are fully ionized, the background which is induced by low-shell electrons in the case of investigation of heavy nuclei using traditional methods will be eliminated.

Concerning the nucleus beam, there are three possible options: RHIC, HERA and LHC. Rough estimations of the parameters of the HERA and LHC based FEL γA colliders are given in [3,4]. Recently, construction of 10 GeV energy electron linac tangentially to the RHİC is discussed in order to investigate lepton-nuclei collisions [5]. If the TESLA-type accelerator [6] is chosen as an electron linac, this will give opportunity to operate e-RHIC as a FEL γA collider, also.

Parameters of $^{232}$Th beams for mentioned options are presented in Table 1, where we use the parameters of gold nucleus for linac-ring option of e-RHIC as reference point [5]. Parameters of TESLA FEL beam [7] are given in Table 2. Luminosity of the γ-nucleus collisions is given by

$$L = \frac{n_\gamma \times n_A}{4\pi \times \sigma_x^{eff} \times \sigma_y^{eff}} \times n_b \times f_{rep}$$

where $n_\gamma$ and $n_A$ are the number of photons and nuclei in the FEL and nucleus bunches, respectively, $n_b$ is the number of electron bunches per train, $f_{rep}$ is TESLA repetition rate, $\sigma_x^{eff}$ and $\sigma_y^{eff}$ are the values which are obtained by choosing the bigger ones of the corresponding horizontal and vertical sizes of the FEL and the nucleus beams (for round beams $\sigma_x = \sigma_y$). Obtained luminosity values are given in the last row of the Table 1 (here we use $n_\gamma = 10^{13}$).

### 3. Excitations of the $^{232}$Th nucleus

Today, more than hundred excitations of the $^{232}$Th nucleus are founded in different experiments [8] and only seven of them are observed in $(\gamma,\gamma')$ reactions. The reasons are following: low statistics, limited energy resolution, neutron induced γ background from electron setup material collisions etc [9]. In Table 3 we present main characteristics of a number of these excitations, namely, all 7 excitations observed in $(\gamma,\gamma')$ reactions and $2^+$ excitations with measured lifetime.

The cross section for the resonant photon scattering is given by the well-known Breit-Wigner formula

$$\sigma(\gamma,\gamma') = \frac{\pi}{E^2} \times \frac{2J_{exc}+1}{2 \times (2J_0+1)} \times \frac{B_{in} B_{out} \Gamma^2}{(E-E_R)^2 + \Gamma^2/4}$$

where E is the c.m. energy of the incoming photon (in our case it is very close to that in the rest frame of the nucleus), $J_{exc}$ and $J_0$ are spins of the excited and ground state of the nucleus, $B_{in}$ and $B_{out}$ are the branching fractions of the excited nucleus into the entrance and exit channels, respectively, $E_R$ is the energy at the resonance and $\Gamma$ is the total width of the excited nucleus. Corresponding resonant cross section are given in Table 3.

The energy of FEL photons needed for excitation of corresponding $^{232}$Th level can be expressed as (for details, see [3, 4]):

$$\omega_0 = \frac{E_{exc}}{2\gamma_{Th}}$$

where $E_{exc}$ is the energy of corresponding excited level, $\gamma_{Th}$ is the Lorentz factor of the $^{232}$Th nucleus. Corresponding values for needed FEL photon energies are given in the Table 4.

Taking into account the energy spread of FEL and the nucleus beam ($\Delta E_\gamma/E_\gamma = \Delta\omega/\omega_0 = 10^{-4}$ and $\Delta E_A/E_A = 10^{-4}$), the approximate value of averaged cross section has been found to be [3]

$$\sigma_{av} \approx \sigma_{res} \frac{\Gamma}{\Delta E_\gamma}$$

where $\Delta E_\gamma \approx 10^{-4} \times E_{exc}$. Corresponding values for averaged cross sections are presented in the last column of Table 3. Multiplying these values with luminosities of corresponding colliders, we obtain number of events per second which are given in Tables 4. As seen typical numbers of events are of order of millions per second, whereas "traditional" experiments deals with hundreds events per day.

We have mentioned above that one of the handicaps of the known ($\gamma,\gamma'$) experiments is low statistics. For example, 714.25 keV and 1078.7 keV excitations are observed in ($\gamma,\gamma'$) reactions [8] but their decay widths are not determined (see Table 3). FEL $\gamma$ - Nucleus colliders will give opportunity to "measure" unknown decay widths using known ones. Indeed, $\Gamma_{unknown}$ can be estimated using following relation

$$\Gamma_{(1)} \cong \frac{E_\gamma^{(1)}}{E_\gamma^{(2)}} \times \frac{N_{(1)}}{N_{(2)}} \times \frac{\sigma_{res}^{(2)}}{\sigma_{res}^{(1)}} \times \Gamma_{(2)}$$

where index 1 (2) corresponds to level with unknown (known) decay width. Let us use the 49.368 keV level as known one and 100 events per second as observation limit. In this case we will be able to "measure" decay width of the 714.25 (1078.7) keV level at FEL $\gamma$ - RHIC collider, if it exceeds $2.5 \cdot 10^{-7}$ ($2.6 \cdot 10^{-6}$) eV.

### 4. Conclusion

We hope that the huge number of events provided by FEL $\gamma$ - A colliders and pure experimental environment will give opportunity to investigate the most of known ~100 excitations of $^{232}$Th nucleus, which are observed by different experiments, in ($\gamma,\gamma'$) reactions, too, as well as to observe a lot of additional levels.

Table 1. Parameters of $^{232}$Th beams

|  | RHIC | HERA | LHC |
|---|---|---|---|
| Lorentz factor | 103 | 413 | 2894 |
| Normalized emittance, $\pi\cdot\mu m$ | 6 | 6 | 6 |
| Bunch population, $10^9$ | 1.7 | 1.7 | 1.7 |
| Ring circumference, m | 3833 | 6336 | 26659 |
| Bunches per ring | 180 | 180 | 608 |
| Bunch spacing, ns | 71 | 96 | 125 |
| RMS beam size at the IP, $\mu m$ | 60 | 30 | 11 |
| Luminosity of $\gamma$-Nd collisions, $10^{30}$ cm$^{-2}$s$^{-1}$ | 2 | 8.5 | 8.5 |

Table 2. Parameters of TESLA FEL beam

| Linac repetition rate, Hz | 5 |
|---|---|
| Number of bunches per train | 11315 |
| Photon energy range, keV | 0.1÷12 |
| Number of photons per bunch, $10^{12}$ | 1÷100 |
| Photon beam divergence, $\mu rad$ | 1 |
| Photon beam diameter, $\mu m$ | 20 |

Table 3. Main characteristics of some of the $^{232}$Th nucleus excitations

| E, keV | $J^\pi$ | $(\gamma,\gamma')$ | $T_{1/2}$, s | $\Gamma$, eV | $\sigma_{res}$, cm$^2$ | $\sigma_{av}$, cm$^2$ |
|---|---|---|---|---|---|---|
| 49.369 | $2^+$ | + | $1.5\cdot10^{-11}$ | $4.4\cdot10^{-5}$ | $5.02\cdot10^{-18}$ | $4.47\cdot10^{-23}$ |
| 714.25 | $1^-$ | + | - | - | $1.43\cdot10^{-20}$ | - |
| 774.1 | $2^+$ | - | $5.8\cdot10^{-12}$ | $1.13\cdot10^{-4}$ | $2.04\cdot10^{-20}$ | $2.97\cdot10^{-24}$ |
| 785.3 | $2^+$ | - | $2.7\cdot10^{-12}$ | $2.4\cdot10^{-4}$ | $1.98\cdot10^{-20}$ | $6.05\cdot10^{-24}$ |
| 1078.7 | $0^+$ | + | - | - | $2.10\cdot10^{-21}$ | - |
| 1387.2 | $2^+$ | - | $1.4\cdot10^{-12}$ | $4.71\cdot10^{-4}$ | $6.35\cdot10^{-21}$ | $2.15\cdot10^{-26}$ |
| 1554.2 | $2^+$ | - | $2.95\cdot10^{-12}$ | $6.94\cdot10^{-3}$ | $5.06\cdot10^{-21}$ | $2.26\cdot10^{-25}$ |
| 2043 | $1^+$ | + | $8.64\cdot10^{-15}$ | 0.077 | $1.75\cdot10^{-21}$ | $6.59\cdot10^{-25}$ |
| 2248 | $1^+$ | + | $1.8\cdot10^{-14}$ | 0.037 | $1.45\cdot10^{-21}$ | $2.39\cdot10^{-25}$ |
| 2274 | $1^+$ | + | $3.9\cdot10^{-14}$ | 0.017 | $1.41\cdot10^{-21}$ | $1.06\cdot10^{-25}$ |
| 2296 | $1^+$ | + | $2.64\cdot10^{-14}$ | 0.025 | $9.26\cdot10^{-22}$ | $1.01\cdot10^{-25}$ |

Table 4. The $^{232}$Th excitations at the FEL$\gamma\otimes$RHIC collider

| E, keV | RHIC $\omega_0$, eV | RHIC $N_{ev}$/s | HERA $\omega_0$, eV | HERA, LHC $N_{ev}$/s | LHC $\omega_0$, eV |
|---|---|---|---|---|---|
| 49.369 | 238 | 8.94·10$^7$ | 59.6 | 3.80·10$^8$ | 8.52 |
| 714.25 | 3454 | - | 860 | - | 123 |
| 774.1 | 3744 | 5.94·10$^6$ | 936 | 2.52·10$^7$ | 133 |
| 785.3 | 3798 | 1.21·10$^7$ | 949 | 5.14·10$^7$ | 135 |
| 1078.7 | 5217 | - | 1304 | - | 186 |
| 1387.2 | 6709 | 4.30·10$^4$ | 1677 | 1.83·10$^5$ | 239 |
| 1554.2 | 7517 | 4.52·10$^5$ | 1879 | 1.92·10$^6$ | 268 |
| 2043 | 9881 | 1.32·10$^6$ | 2470 | 5.60·10$^6$ | 352 |
| 2248 | 10873 | 4.78·10$^5$ | 2718 | 2.03·10$^6$ | 388 |
| 2274 | 10999 | 2.12·10$^5$ | 2749 | 9.01·10$^5$ | 392 |
| 2296 | 11105 | 2.02·10$^5$ | 2776 | 8.58·10$^5$ | 396 |